\documentclass[journal]{IEEEtran}
\usepackage{amsmath,amsfonts}
\usepackage{algorithmic}
\usepackage{array}

\usepackage{textcomp}
\usepackage{stfloats}
\usepackage{url}
\usepackage{verbatim}
\usepackage[pdftex]{graphicx}
\graphicspath{{../pdf/}{../jpeg/}}
\DeclareGraphicsExtensions{.pdf,.jpeg,.png,.tif}

\usepackage{xurl}
\usepackage{hyperref}
\usepackage{cleveref}
\usepackage{amsmath,amssymb,amsfonts}
\usepackage{mdwmath}
\usepackage{mdwtab}
\usepackage{eqparbox}
\usepackage{nccmath}
\usepackage{cite}
\usepackage{multirow}
\usepackage[]{graphicx}
\usepackage{xcolor,soul,framed} 
\usepackage{color}
\usepackage{amsthm}
\newtheorem{remark}{Remark}
\usepackage{threeparttable} 

\usepackage{caption}
\usepackage[labelformat=simple]{subcaption}

\usepackage{float}
\usepackage{verbatim}
\usepackage{bm}
\usepackage{xcolor}

\usepackage{lccaps}
\usepackage{optidef}
\usepackage{textcomp}
\usepackage{makecell}
\usepackage{stackengine}
\usepackage{bm}

\def\BibTeX{{\rm B\kern-.05em{\sc i\kern-.025em b}\kern-.08em
    T\kern-.1667em\lower.7ex\hbox{E}\kern-.125emX}}
\usepackage{balance}

\IEEEoverridecommandlockouts

\begin{document}

\title{Optimal Day-Ahead Scheduling of Fast EV Charging Station
With Multi-Stage Battery Degradation Model}

\author{
Yihao Wan,~\IEEEmembership{Student Member,~IEEE,}
Daniel Gebbran,~\IEEEmembership{Member,~IEEE,}
Ramadhani Kurniawan Subroto,~\IEEEmembership{Member,~IEEE,} and
Tomislav Dragičević,~\IEEEmembership{Senior Member,~IEEE}
\vspace{-1.6em}
\thanks{Yihao Wan, Ramadhani Kurniawan Subroto, and Tomislav
Dragičević are with the Department of Electrical Engineering,
Technical University of Denmark, Copenhagen, Denmark. Daniel Gebbran
is with Equilibrium Energy, California, USA (e-mails:
wanyh@dtu.dk, daniel.gebbran@equilibriumenergy.com,
rkusu@dtu.dk, tomdr@dtu.dk).}
\thanks{\copyright~2023 IEEE. Personal use of this material is
permitted. Permission from IEEE must be obtained for all other uses,
in any current or future media, including reprinting/republishing
this material for advertising or promotional purposes, creating new
collective works, for resale or redistribution to servers or lists,
or reuse of any copyrighted component of this work in other works.
This is the author-accepted manuscript of Y. Wan, D. Gebbran,
R. K. Subroto, and T. Dragičević, ``Optimal Day-Ahead Scheduling
of Fast EV Charging Station With Multi-Stage Battery Degradation
Model,'' \textit{IEEE Transactions on Energy Conversion}, vol. 39,
no. 2, pp. 872--883, Jun. 2024,
doi: 10.1109/TEC.2023.3335661.}
}

\maketitle

\begin{abstract}
The paper proposes a day-ahead scheduling framework with a novel multi-stage battery degradation modeling method for an electric vehicle (EV) fast charging station (FCS) equipped with a battery energy storage system (BESS). Unlike previous studies, which employ a single battery degradation model to represent the aging process, this paper proposes a novel multi-stage battery degradation modeling method to capture the degradation process across the whole lifespan accurately. Subsequently, the multi-stage model is explicitly integrated into the proposed adaptive optimization framework in a computationally tractable way, thus having important practical implications in the field. The paper provides case studies to demonstrate the effectiveness of the proposed modeling method on a selected cycle aging model in reducing the operation cost of FCS with BESS operating in different stages. As a result, the overall operation cost with the multi-stage model is around 2.9$\%$ on average lower than the single-stage model counterpart. In addition, results show that with the increasing number of divided stages, the model error decreases and becomes stable, while the reduced operation cost compared with the single-stage model increases and saturates. Finally, we apply the multi-stage framework considering other conventional degradation models to show the superiority of the proposed method.
\end{abstract}


\section{Introduction}
\IEEEPARstart{W}{ith} the increasing transportation electrification to reduce Greenhouse Gas (GHG) emissions, electric vehicles (EVs) are gaining a significant market share \cite{outlook2021accelerating}. Although EVs are becoming more popular, one of the major bottlenecks for the large-scale replacement of Internal Combustion Engine (ICE) vehicles is the lack of fast-charging infrastructures, particularly on the highways between cities and rural districts. Fast charging stations (FCSs) are important infrastructures that can provide a quick recharging experience for EVs similar to the experience of refueling ICE vehicles \cite{morrissey2016future}. However, due to the impulsive nature of the load profile in FCSs, they may cause adverse impacts on the stability of the grid, especially for those installed in rural areas with weak grid connections \cite{mahfouz2019grid,khalid2019comprehensive}. In addition, the installation of FCSs may require upgrading the electrical infrastructure (e.g., installation of new transformers and power lines), resulting in high installation costs for both the charge point operator (CPO) and the distribution system operator (DSO), who usually transfers its share of the costs to the CPO in the form of grid-connection fee \cite{rafi2020comprehensive}. 

To partially mitigate the above issues, battery energy storage systems (BESSs) can be integrated into FCSs, acting as a buffer between the grid and the EVs \cite{zheng2018energy, energydk2022}. BESSs can not only partially mitigate the high cost involving grid connection fees but can also provide premium grid services such as frequency containment reserve (FCR) \cite{rafi2020comprehensive}. In addition, the BESS can help reduce the utility charges for the FCS by doing energy price arbitrage \cite{yang2021comprehensive}. 

One of the most commonly researched applications of BESSs is energy arbitrage (i.e., charging at low prices during off-peak intervals and selling the stored energy during the peak load at high price intervals), where the operators take advantage of the electricity spot price differences. The key challenge with arbitrage is obtaining an economic dispatch strategy to achieve a trade-off among the arbitrage revenue, battery degradation, and energy consumption costs for minimal operation cost. Therefore, battery usage should also be explicitly considered in the optimization problem because more usage implies more losses and degradation \cite{wankmuller2017impact}. 

\subsection{Background}
To formulate the battery degradation cost in the scheduling framework, the simplest way is introducing operational factors of battery into the objective function or constraints such as the number of cycles \cite{alramlawi2020design}, state-of-charge (SoC) \cite{8259445}, depth-of-discharge (DoD) \cite{9340236}, current rate \cite{7177105}, charging/discharging power \cite{elkazaz2020energy, fan2021whole}, etc. Although those methods make the scheduling problem possible to be solved with common mathematical programming methods such as linear programming (LP) or quadratic programming (QP). However, since battery degradation behavior is highly non-linear \cite{maheshwari2020optimizing}, these terms do not explicitly represent the battery capacity loss throughout its operational lifetime, which may lead to overuse or underuse of the battery. In addition, the optimal scheduling is highly dependent on the selected weighting coefficients for the introduced operational factors, which are not straightforward to select \cite{faraji2020optimization,merabet2022energy}. Therefore, though using the simple term to account for the battery usage reduces the computational complexity of the operational planning problem, it doesn't explicitly represent the actual battery degradation which would result in suboptimal operations and incur higher costs in the long term.

To accurately account for battery usage in scheduling problems, different degradation models have been previously proposed. In our previous work\cite{wan2022optimal}, a DoD-based degradation model is implemented by discretizing the SoC of the battery during operation, which reduced the small charge/discharge cycles, thus reducing the battery capacity loss while maintaining near optimal revenue from grid services. However, the battery charging/discharging current also contributes to the degradation, and the same cycle depth for low current and high current could be induced if the charge time is sufficient. In \cite{wang2020impact}, a combined factor-based battery degradation model based on different stress-based models is proposed, while the model accuracy is only validated in the early aging stage. The optimal operation of BESS based on a multi-factor battery cycling model is proposed in \cite{abdulla2018optimal}, which employs the rainflow counting algorithm (RCA) for capacity loss estimation. RCA algorithm is widely used in the context of fatigue analysis and damage estimation \cite{barragan2022enhancement}. However, due to the lack of an analytical formula for RCA, it is hard to implement it explicitly in an optimization problem \cite{xu2017factoring}. Different ways of solving this are proposed, such as using problem-specific solvers for the nonlinear RCA term \cite{schneider2020rechargeable}, approximation \cite{lee2022novel,ke2014control}, etc. 

In any case, all the aforementioned methods are based on a single-stage battery degradation model, considering only a single degradation manner across the whole battery lifetime, which is usually not realistic due to the nonlinear and complex battery degradation process. The degradation behavior under various stressing conditions may experience multiple patterns over the lifespan \cite{gao2017lithium}.

Given the complex aging mechanism across the lifespan under diverse stress situations, the degradation pattern is constant within a single stage but varies among other stages. Therefore, a single battery degradation model cannot represent the dynamic and complex aging process over the entire battery lifetime. Some attempt to model the battery aging process in different degradation stages. The multi-stage battery degradation modeling method in \cite{gao2017lithium} describes the varying charging stress at different battery states. However, it is not employed in practical scheduling problems, and the model also lacks flexibility or adaptability for more stress variables. In \cite{qin2022transferable}, a multi-stage model with cycling discrepancy learning for SoH estimation is proposed. It estimates the degrading trend in different stages. Similarly, a one-shot battery degradation trajectory prediction method based on deep learning is proposed \cite{li2021one}. However, the two methods are used to project the cycle life, and the stress conditions are not considered. Thus, they cannot be applied explicitly to operational planning problems. In \cite{liu2022capacity}, the aging mechanism and SoH prediction across the total lifespan are investigated based on three stages with a neural network (NN). However, the model is developed based on electrochemical impedance spectroscopy (EIS), which cannot be directly employed due to the unavailability of detailed cell conditions during the real-time planning stages. Moreover, the scheduling problem with an NN-based degradation model is highly nonlinear and nonconvex, which makes it difficult to solve. Another work in \cite{xiong2018lithium} proposes to estimate the battery SoH and remaining useful life (RUL) at different aging states with a linear aging model based on a moving window. Nonetheless, it depends on the extraction and fitting of the health indicator. The model is not explicitly related to the stress variables but to the cycling patterns under specific conditions, which is not applicable to operational planning problems. In summary, the works above reveal that battery degradation patterns change in different battery states. Though they are proposed to estimate the battery SoH or RUL at different stages, most models are not feasible for operational planning problems as the battery degradation is not explicitly formulated with different stress variables.

\subsection{Literature gap}
In general, the above-mentioned methods utilize a single-stage battery degradation model for the scheduling problem throughout the battery lifespan. The degradation pattern within a single stage would change and differ from other stages due to the complicated and coupling aging mechanism, resulting in a nonlinear battery aging process. As a result, no single model can accurately capture battery degradation across its entire lifespan. In addition, the scheduling strategy should also be adaptively changed according to the varying degradation patterns within different operating stages. Operating a BESS without acknowledging this fact results in suboptimal operations, leading to an increase in overall operation costs. 
\begin{figure}
	\begin{center}
		\includegraphics[width=0.9\linewidth]{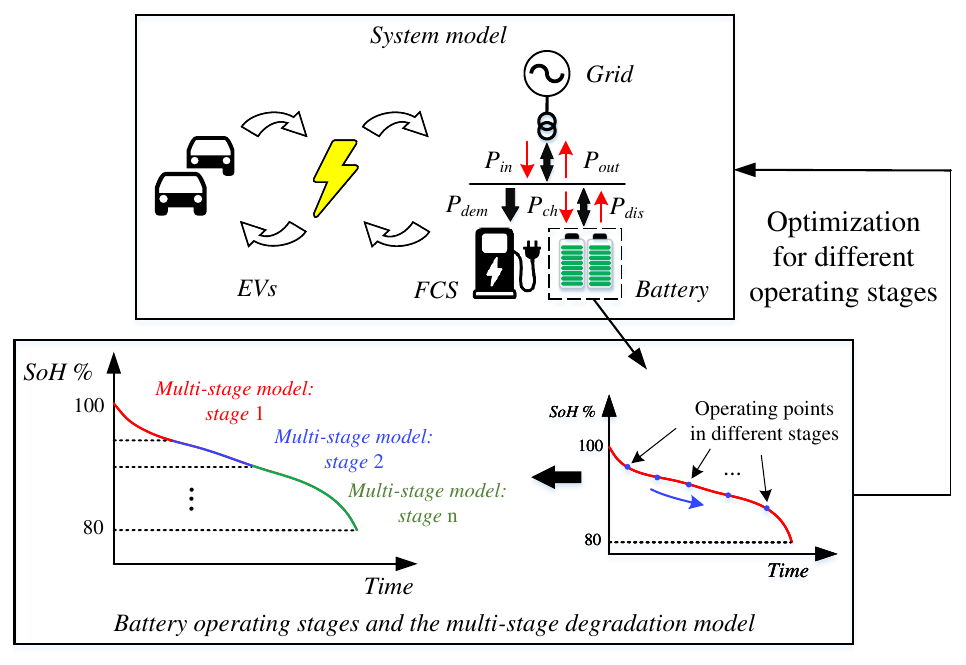}\\
		\caption{Scheduling framework with the proposed multi-stage degradation model.}
        \vspace{-1.2em}
		\label{Framework}
	\end{center}
\end{figure}

\subsection{Contributions}
This paper aims to bridge the aforementioned gaps by proposing a multi-stage battery degradation modeling method and integrating the multi-stage models explicitly into an optimization framework, as shown in Fig. \ref{Framework}. In particular, we focus on the day-ahead scheduling optimization with the multi-stage battery degradation model, where the battery cycle degradation process should be properly modeled. The proposed multi-stage modeling method is demonstrated on a generic cycle aging model selected in \cite{motapon2020generic}. {red}{Different stressing factors} are considered in the degradation model, developed respectively for different operational stages. The models are then adaptively implemented in the scheduling model with respect to the battery operating stages. The main contributions of the paper are summarized below:
\begin{itemize}
\item A novel multi-stage battery degradation modeling method. The proposed multi-stage degradation modeling method can accurately project the nonlinear battery degradation process throughout its lifespan by capturing the varying aging patterns. In particular, by segmenting the battery lifetime into several stages according to the SoH ranges, the degradation model considering {red}{the} stress factors is developed for each stage, respectively. {red}{The proposed approach is applicable to various conventional single-stage battery degradation models.}
\item An adaptive scheduling framework with the multi-stage degradation model. A novel optimization framework {red}{for the day-ahead scheduling problem} is proposed {red}{by} explicitly {red}{accounting} for the multi-stage battery degradation model in a computationally tractable way for real-time implementation. The multi-stage model is adaptively integrated into the scheduling model at different battery stages. The effectiveness of the proposed optimization framework is demonstrated with case studies at different stages over the lifetime, while the computational burden is explicitly measured to prove the feasibility of real-time implementation.
\item Analysis regarding the number of divided stages and validations on various models. The performance of the proposed multi-stage modeling method is further investigated by increasing the number of divided stages. The {red}{reduced} operation cost with the proposed optimization framework based on different multi-stage degradation models {red}{regarding the single-stage model counterpart} is compared. For further validations, the multi-stage framework is applied to other conventional {red}{single-stage} degradation model and {red}{compared with a multi-stage modeling approach to} show the superiority of the proposed method.
\end{itemize}

The rest of the paper is structured as follows. Section \uppercase\expandafter{\romannumeral2} introduces the proposed multi-stage modeling concept based on a specific degradation model. Section \uppercase\expandafter{\romannumeral3} introduces the mathematical modeling of the FCS with a BESS, combining the scheduling model of the system with the proposed multi-stage battery degradation model. Section \uppercase\expandafter{\romannumeral4} compares the scheduling results of the single- and multi-stage battery degradation model based on the case studies of example days for different operating stages. Section \uppercase\expandafter{\romannumeral5} concludes the paper.

\section{Modeling of battery degradation}
To formulate the aging cost for the scheduling problems, the battery degradation process consisting of calendar and cycle aging should be properly modeled. However, the study does not consider battery calendar aging due to its independence from operations and negligible influence on short-term operational planning problems. Therefore, the battery cycle degradation model is developed for evaluating the capacity loss during operations in this section. To demonstrate the proposed multi-stage degradation modeling framework, a generic cycle life model for lithium-ion batteries from \cite{motapon2020generic} is employed as the fundamental model. In addition, to identify the parameters of the models, battery experimental cycling test data in \cite{batteryarchive} is utilized, which provides battery cycle aging test data under various conditions (e.g., different DoD levels, current rates, {red}{and temperatures}).

\subsection{Single-stage battery degradation model}
The degradation model considers {red}{multiple} stress factors, DoD, current rate {red}{and temperature}. Moreover, charging and discharging currents are assumed to contribute equally to capacity loss. {red}{The temperature stress is formulated with Arrhenius equation \cite{smith2012comparison}. In addition,} according to Wöhler's approximation and Miner's rule, the battery degradation per cycle is calculated as the stress amplitude subjected to the peak stress the material can sustain. By analogy, based on the assumption of the same degrading contribution for charging and discharging current, the combined stress factor model is formulated as \cite{smith2012comparison,laresgoiti2015modeling}
\begin{equation}
\small
\theta(t) = \theta_{DoD}(t) \cdot \theta_{I_{ch/dis}}(t) \cdot {red} \theta_{T}(t),
\label{degrading rate}
\end{equation}

\begin{equation}
\small
\theta_{DoD}(t) = (\frac{DoD(t)}{DoD_{ref}})^{\frac{1}{\alpha}}, 
\label{DOD rate}
\end{equation}

\begin{equation}
\small
\theta_{I_{ch/dis}}(t) = (\frac{I_{ch/dis}(t)}{I_{ref}})^{\frac{1}{\beta}}, 
\label{current rate}
\end{equation}

{red}{
\begin{equation}
\small
\theta_{T}(t) = exp\left[-\psi\left(\frac{1}{T_a(t)}-\frac{1}{T_{ref}}\right)\right]
\label{Temp}
\end{equation}}

\noindent where $\theta$ represents the combined stressing factor model, $\alpha$ and $\beta$ are the stress exponents for DoD and current rate, respectively. {red}{$\psi$ is the Arrhenius rate constant}, $DoD_{ref}$ and $I_{ref}$ are the peak stress amplitude for the DoD and battery reference current, {red}{$T_a$ and $T_{ref}$ are the ambient and reference temperature respectively.}

The maximum number of cycles to EoL regarding the stress condition is expressed as
\begin{equation}
\begin{aligned}
\small
N_c(t) = \frac{N_{cref}}{\theta(t)} =& N_{cref} \cdot \left(\frac{DoD(t)}{DoD_{ref}}\right)^{-\frac{1}{\alpha}} \cdot \left(\frac{I_{ch/dis}(t)}{I_{ref}}\right)^{-\frac{1}{\beta}} \\& {red}\cdot exp\left[-\psi\left(\frac{1}{T_{ref}}-\frac{1}{T_a(t)}\right)\right],
\label{cycle_number}
\end{aligned}
\end{equation}
\noindent where $N_{cref}$ is the number of cycles to EoL for the reference cycle condition at $DoD_{ref}$ and $I_{ref}$, $N_c$ is the number of cycles to EoL for the resultant stress factors.

The battery capacity degradation over its lifetime, considering the non-linear aging characteristics, is expressed as
\begin{equation}
\small
Q(t) = Q_{BoL} - \varphi(t)^{\xi} \cdot (Q_{BoL} - Q_{EoL}),
\label{Capacity_process}
\end{equation}
\noindent where $Q$ denotes the battery capacity, $Q_{BoL}$ and $Q_{EoL}$ denote the battery capacity at BoL and EoL, respectively. $Q_{EoL}$ is assumed to be 80$\%$ of $Q_{BoL}$. $\varphi$ is the aging index, representing the contribution of the cycles to the aging of the battery, while $\xi$ is the aging exponent of capacity, expressed below.
\begin{equation}
\small
\xi = \frac{ln(\frac{Q_{BoL}-Q_5}{Q_{BoL}-Q_{EoL}})}{ln(\frac{N_{ref}}{N_{cref}})},
\label{xi}
\end{equation}
\noindent where $Q_5$ denotes 5$\%$ capacity losses of $Q_{BoL}$, $N_{ref}$ is the number of cycles to $Q_5$ for reference testing condition.

The model parameters are estimated below.
\begin{equation}
\small
N_{ci} = \frac{N_{cref}\cdot N_i}{N_{ref}},
\label{cycles}
\end{equation}

\noindent where $N_{ci}$ and $N_i$ denote the number of cycles to EoL and $Q_5$ in cycling test condition $i\in \{2,..., n\}$ for parameters identification. The stress exponents are formulated below.

\begin{equation}
\small
\alpha = -\frac{ln(\frac{DoD_i}{DoD_{ref}})}{ln(\frac{N_{ci}}{N_{cref}})},
\label{alpha}
\end{equation}

\begin{equation}
\small
\beta = -\frac{ln(\frac{I_i}{I_{ref}})}{ln(\frac{N_{ci}}{N_{cref}})},
\label{beta}
\end{equation}

\begin{equation}
\small
{red}\psi = \frac{ln(\frac{N_{ci}}{N_{cj}})}{\frac{1}{T_i}-\frac{1}{T_{ref}}}.
\label{arr_const}
\end{equation} 
The test conditions for identifying the model parameters are listed in Table \uppercase\expandafter{\romannumeral1}. Finally, we assign the identified parameters from Table \uppercase\expandafter{\romannumeral2} into (\ref{degrading rate})-(\ref{current rate}), and the single-stage battery degradation model is obtained. {red}{It is noteworthy that while the combined stress model is parameterized with experimental data under specific cycling conditions, the formulated model can still yield reasonable evaluation accuracy of battery degradation under different stressing conditions by providing general trends of battery degradation \cite{motapon2020generic}. }
\bgroup
\def\arraystretch{1.5}
\begin{table}[h]
\caption{Cycling conditions for parameter identification.\label{tab:tests}}
\centering
    \begin{threeparttable}
        \begin{tabular}{c c c c c} 
        \hline
        Conditions & $1^*$ & 2 & 3 & 4\\
        \hline
        DoD & 100$\%$ & 60$\%$ & 100$\%$ &{red}100$\%$\\
        C-rate & 0.5C & 0.5C& 2C&{red}2C\\
        {red}Temperature & {red}25\textcelsius & {red}25\textcelsius& {red}25\textcelsius& {red}35\textcelsius\\
        \hline
        \end{tabular}
    \end{threeparttable}
    \begin{tablenotes}
        \footnotesize
        \item \quad \quad \quad \ {red}* Reference condition
    \end{tablenotes}
\end{table}
\egroup

\bgroup
\def\arraystretch{1.5}
\begin{table}[h]
\caption{Parameter identification for single-stage battery degradation model.\label{tab:single}}
\centering
\begin{tabular}{c c c c c}
\hline
\multicolumn{5}{c}{Inputs for parameters identification}\\
\hline
$N_1$ & $N_2$ & $N_3$ & {red}$N_4$ & $N_{c1}$\\
22 & 256 & 18 & {red}20 & 513\\
\hline
\multicolumn{5}{c}{Model parameters identification}\\
\hline
$N_{cref}$ & $\alpha$ & $\beta$ & {red}$\psi$ & $\xi$\\
513 & 0.2043 & 6.9083 &{red}-9.2190 & 0.4402\\
\hline
\end{tabular}
\end{table}
\egroup

\subsection{Multi-stage battery degradation model}
The multi-stage battery degradation model is developed based on the experimental data of battery degradation throughout its lifetime. In particular, as the degrading rate changes adaptively in different states, the multi-stage degradation modeling method allows the model to capture the different degradation patterns under various stress conditions within different SoH ranges of BESS. This effect is depicted in the bottom left image in Fig. \ref{Framework}. In this paper, the battery degrading process until it reaches EoL with 80$\%$ SoH is evenly divided into three stages (i.e., stage 1: 100$\%$-93.3$\%$, stage 2: 93.3$\%$-86.6$\%$ and stage 3: 86.6$\%$-80$\%$ SoH), where the battery degradation model is parameterized respectively. More divided stages increase the accuracy of the degradation model and more economical operations. The performance of the model with the different number of divided stages will be discussed in the following sections in terms of operation costs with the scheduling results and model estimation error. 

In particular, to identify the model parameters for each stage, in the same way presented in Section \uppercase\expandafter{\romannumeral2}.A, the experimental test data corresponding to each stage (i.e., different SoH range), is employed \cite{preger2020degradation}. The model parameters for each stage are listed in Table \ref{tab:stage1}-\ref{tab:stage3}.
\bgroup
\def\arraystretch{1.5}
\begin{table}[H]
\caption{Multi-stage battery degradation model: Stage 1.\label{tab:stage1}}
\centering
\begin{tabular}{c c c c c}
\hline
\multicolumn{5}{c}{Inputs for parameters identification}\\
\hline
$N_1$ & $N_2$ & $N_3$ & {red} $N_4$ & $N_{c1}$\\
36 & 378 & 34 & {red}39& 513\\
\hline
\multicolumn{5}{c}{Model parameters identification}\\
\hline
$N_{cref}$ & $\alpha$ & $\beta$ & {red}$\psi$ & $\xi$\\
513 & 0.2172 & 24.2535 & {red}-163.7827& 0.3952\\
\hline
\end{tabular}
\end{table}
\egroup

\bgroup
\def\arraystretch{1.5}
\begin{table}[H]
\caption{Multi-stage battery degradation model: Stage 2.\label{tab:stage2}}
\centering
\begin{tabular}{c c c c c}
\hline
\multicolumn{5}{c}{Inputs for parameters identification}\\
\hline
$N_1$ & $N_2$ & $N_3$ & {red} $N_4$ & $N_{c1}$\\
231 & 1586 & 201 & {red}280 & 513\\
\hline
\multicolumn{5}{c}{Model parameters identification}\\
\hline
$N_{cref}$ & $\alpha$ & $\beta$& {red}$\psi$ & $\xi$\\
513 & 0.2652 & 9.9653 &{red}-50.6226 &0.4470\\
\hline
\end{tabular}
\end{table}
\egroup

\bgroup
\def\arraystretch{1.5}
\begin{table}[H]
\caption{Multi-stage battery degradation model: Stage 3.\label{tab:stage3}}
\centering
\begin{tabular}{c c c c c}
\hline
\multicolumn{5}{c}{Inputs for parameters identification}\\
\hline
$N_1$ & $N_2$ & $N_3$& {red} $N_4$ & $N_{c1}$\\
513 & 3628 & 562 & {red}771& 513\\
\hline
\multicolumn{5}{c}{Model parameters identification}\\
\hline
$N_{cref}$ & $\alpha$ & $\beta$& {red}$\psi$ & $\xi$\\
513 & 0.2611 & -15.1963 &{red}-27.6663 &0.5066\\
\hline
\end{tabular}
\end{table}
\egroup

To validate the performance of the proposed multi-stage battery degradation model, additional cycling test datasets that are different from those used for parameter identification are employed. Table \ref{pef_com} shows the root-mean-square error (RMSE) of the two models based on the experimental data during different battery operating stages when the battery is cycled at 60$\%$ DoD, 3C current rate and 25 \textcelsius, and {red}{100$\%$ DoD, 1C current rate and 15 \textcelsius, respectively}. Despite the close performance between the two models in the initial degradation stage 1, the multi-stage battery degradation model outperforms the single battery degradation model in terms of estimation accuracy for all the stages across the battery lifespan. 
\bgroup
\def\arraystretch{1.5}
\begin{table}[H]
\caption{Model performance under different tests at different stages.\label{pef_com}}
\centering
\begin{tabular}{c c c c}
\hline
Tests & Stages & Single-stage model & Multi-stage model\\
\hline
\multirow{3}{*}{Test {red}{5}}& Stage 1 & 0.78$\%$ & 0.50$\%$ \\
& Stage 2 & 2.21$\%$ & 0.42$\%$ \\
& Stage 3 & 3.15$\%$ & 1.10$\%$ \\
\hline
\multirow{3}{*}{Test {red}{6}}& Stage 1 & {red}1.00$\%$ & {red}0.78$\%$  \\
& Stage 2 & {red}2.37$\%$ & {red}0.75$\%$ \\
& Stage 3 & {red}2.86$\%$ & {red}1.28$\%$ \\
\hline
\end{tabular}
\end{table}
\egroup

\begin{remark}
The single-stage degradation model only accounts for a single degradation pattern, which differs between different stages, resulting in increased degradation estimation error. The multi-stage model is built based on the experimental data of battery aging from BoL to EoL by separating the lifespan into different stages. Therefore, the multi-stage model could be considered a piecewise linearization of the real battery degradation process. In this way, the multi-stage battery degradation model can capture multiple degradation patterns at different stages so that the model accuracy is improved. Afterward, the models built for each stage could be adaptively integrated into the scheduling model to achieve economic dispatch strategies throughout the battery lifetime, which will be elaborated on in the following sections.\end{remark}

\section{System modeling and problem formulation}
In this work, we have considered a scheduling model that determines the day-ahead optimal dispatch strategy for the FCS system. The demand profile from the electric vehicles is obtained through load prediction, and the electricity prices are known in advance from the supplier. Battery usage is taken into consideration in the scheduling model by incorporating the battery degradation model as a penalty term, with a cost coefficient converted into a battery usage cost. {red}{Due to the improvement of heating, ventilation, and air conditioning systems for the stationary battery system \cite{lin2023optimized}, the temperature can be independently controlled regardless of the operational planning. Therefore, the cell temperature is maintained at reference temperature 25 $^{\circ}$C with the battery thermal management system.} The optimal scheduling strategy for specific loads and prices during a day for balancing the energy arbitrage revenue, battery usage, and energy consumption cost could be obtained. The formulation of the optimization problem is as below.

\subsection{Objective function}
The objective function (\ref{objective}) is formulated to optimize the dispatch strategies, $P^{ch}_t$ and $P^{dis}_t$ at each discrete time step $t$, within a finite time horizon (i.e., $t \in \mathcal{T} = \{1, 2,..., T\}$) for minimum daily operation cost, achieving the best trade-off among energy arbitrage, load supply and battery degradation cost. In equation (\ref{objective}), the first term yields the power supply cost, which could be minimized by exploiting temporal price differentials on the electricity spot market prices. Meanwhile, the second term accounts for the cost associated with battery degradation. The objective function which reflects this is expressed below
\begin{equation}
\small
\underset{P^{ch}_t, P^{dis}_t}{min} \sum^T_{t=1}(P^{in}_t - P^{out}_t) \cdot p_t + \lambda_k \cdot C_t,
\label{objective}
\end{equation}

\noindent where $P^{in}_t$ and $P^{out}_t$ are the energy flow between the grid and FCS, $P^{ch}_t$ and $P^{dis}_t$ are the charging and discharging power commands for the BESS, $p_t$ is the time-of-use (ToU) electricity spot market price, $C_t$ denotes the battery capacity losses, and a cost coefficient $\lambda_k$ is assigned. In addition, the cost coefficient $\lambda_k$ represents the replacement cost of the battery, assigned with a constant value across the battery's operational lifespan. Therefore, the objective function aims to minimize the overall operation cost of the system while considering the battery degradation. 

The scheduling problem is solved concerning the battery charging power $P^{ch}_t$ and discharging power $P^{dis}_t$, the energy flow $P^{in}_t$ (grid to the FCS) and $ P^{out}_t$ (FCS to the grid) between the grid and system, the operation strategy of the battery $DoD_t$ and $I_t^{ch/dis}$, and the resultant capacity losses $C_t$. The distinction between the power terms is also depicted in Fig. \ref{Framework}.

\subsection{Operation constraints}
\subsubsection{Energy balance}
The energy balance is achieved by utilizing the BESS as a buffer between the grid and FCS, which is formulated as
\begin{equation}
\small
P^{in}_t - P^{out}_t = P^{ch}_t - P^{dis}_t + P_{dem},
\label{energy balance}
\end{equation}
\noindent where $P_{dem}$ denotes the load demand of the vehicles.

\subsubsection{Battery operation}
The evolution of the battery during the operation is formulated in (\ref{SOC_operation}). In addition, to ensure the SoC at the end of the time horizon equals an expected value $SoC_{end}$, constraint (\ref{ini_end}) is introduced. Moreover, the SoC has a safe battery operating range. Based on the above analysis, the battery operation constraints are shown below
\begin{equation}
\small
SoC_t = SoC_{t-1} + \frac{\tau}{E_{bat}}(\eta_{ch} \cdot P^{ch}_t - \frac{P^{dis}_t}{\eta_{dis}}),
\label{SOC_operation}
\end{equation}

\begin{equation}
\small
SoC_{t=T} = SoC_{end}, 
\label{ini_end}
\end{equation}

\begin{equation}
\small
SoC_{min}\leq SoC_t \leq SoC_{max}, 
\label{soc_lim}
\end{equation}
\noindent where $SoC_t$ denotes the state-of-charge (SoC) of battery at time $t$, $\tau$ denotes the time step, $E_{bat}$ is the capacity of battery, $\eta_{ch}$ and $\eta_{dis}$ are the charging and discharging efficiency respectively, $SoC_{min}$ and $SoC_{max}$ are the SoC range for battery operation.

\subsubsection{Power limits}
During the operation, the charging and discharging power is limited within the battery power ratings, as shown below:
\begin{equation}
\small
0\leq P^{ch}_t \leq P_{max}, 
\label{pch_m}
\end{equation}

\begin{equation}
\small
0\leq P^{dis}_t \leq P_{max}, 
\label{pdis_m}
\end{equation}
\noindent where $P_{max}$ is the maximum battery charging/discharging power.

In addition, the power transmission between the FCS and the grid should be restricted to ensure stability 

\begin{equation}
\small
-P_{gmax}\leq P^{in}_t - P^{out}_t\leq P_{gmax}, 
\label{pgrid_lim}
\end{equation}
\noindent where $P_{gmax}$ is maximum energy flow into the FCS.

\subsection{Cost for battery degradation}
The real-time implementation of the degradation model is important to {red}{consider battery degradation in the scheduling problem explicitly}. The aging index represents the contribution of the cycles to the aging of the battery. As the battery degradation model could be considered discretized over a finite time horizon, the battery degradation could be formulated by the cumulative aging index over a certain period at different DoD levels and current rates, which is expressed below

\begin{equation}
\begin{aligned}
\small
Q_{loss} = \sum^T_{t=1} \frac{\theta_t}{N_{cref}} = &\sum^T_{t=1} \frac{1}{N_{cref}} \cdot (\frac{DoD_t}{DoD_{ref}})^{\frac{1}{\alpha}} \cdot (\frac{I^{ch/dis}_t}{I_{ref}})^{\frac{1}{\beta}} \cdot \\& {red}\cdot exp\left[-\psi\left(\frac{1}{T_a}-\frac{1}{T_{ref}}\right)\right],
\label{battery cost}
\end{aligned}
\end{equation}

In addition, an incomplete half cycle or full cycle may happen in each time step and the whole future optimization window, as the $DoD_{ref}$ = 100$\%$, $I_{ref} = 0.5C$, and {red}{$T_a = T_{ref} = 25$\textcelsius}, based on the approximation in \cite{tran2013energy}, replacing the $C_t$ in the objective function (\ref{objective}) with the battery degrading equation as below

\begin{equation}
\small
C_t = \sum^T_{t=1} \frac{(\triangle DoD_t)^{\frac{1}{\alpha}} \cdot (2 I^{ch/dis}_t)^{\frac{1}{\beta}}}{N_{cref}}.
\label{battery cost}
\end{equation}

In general, by replacing the battery usage term $C_t$ in equation (\ref{battery cost}) with the corresponding model parameters, the scheduling model {red}{incorporating} the single-stage and multi-stage degradation models are built respectively \footnote{We use a linear approximation for calculating the DoD term as we did before \cite{wan2022optimal}. Incorporating the RCA for cycle depth $\triangle DoD$ in an optimization problem \cite{lee2022novel,xu2017factoring} is still an open research question, and our approach is similar to other works employing a DoD linearization method \cite{chen2021bargaining,liu2017economic,qin2018stochastic}.}. 

\begin{remark}
The proposed multi-stage degradation modeling method is independent of various empirical/semi-empirical degradation models. The proposed optimization framework with the proposed multi-stage battery degradation model concept can still be employed in the same way for other battery degradation models, which could be integrated to improve the economic dispatch strategies throughout the battery lifespan.
\end{remark}

\section{Case studies}
\begin{figure}
	\begin{center}
		\includegraphics[width=0.9\linewidth]{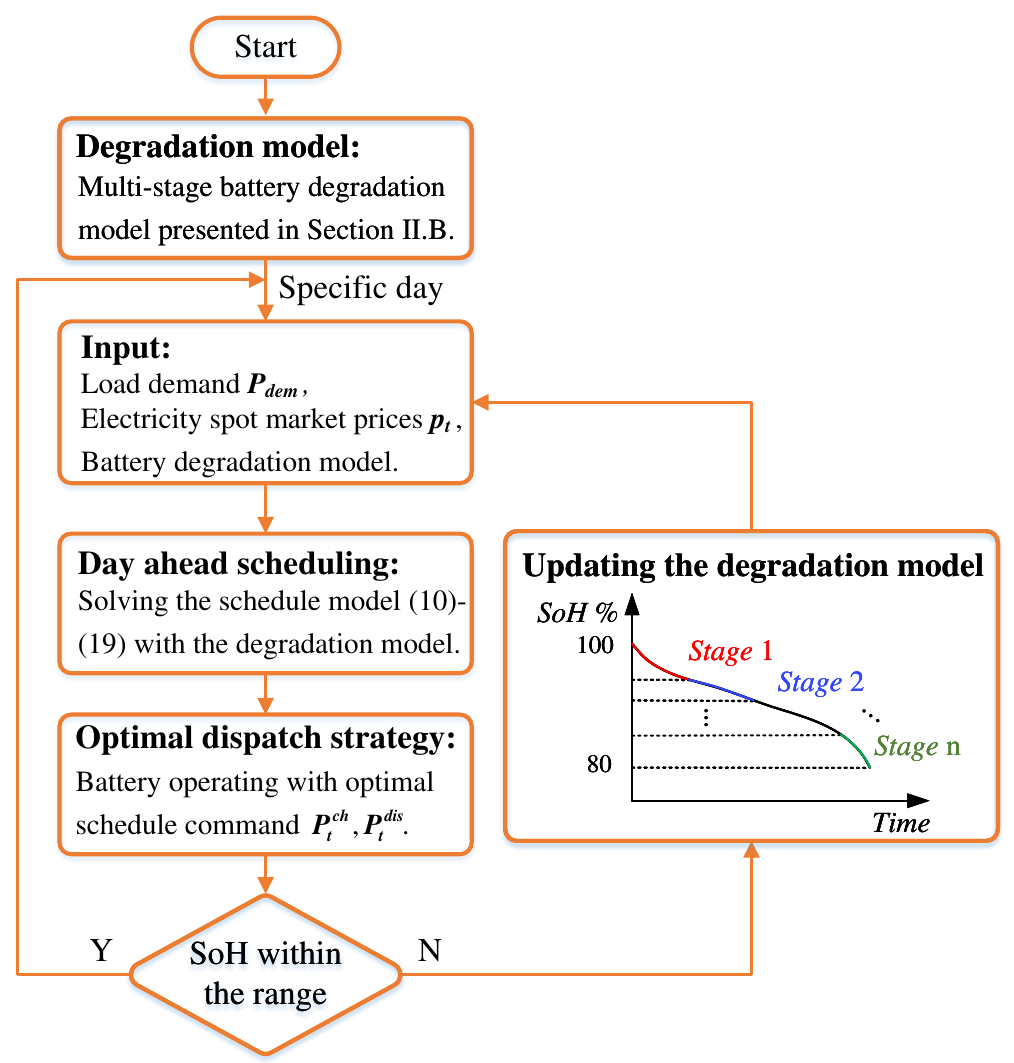}\\
		\caption{Flowchart of the proposed adaptive optimization framework with the multi-stage battery degradation model.}
		\label{Flowchart}
	\end{center}
\end{figure}

To validate the performance of the proposed multi-stage battery degradation model, the proposed framework is implemented in an FCS operation problem and compared with the conventional single-stage battery degradation counterparts in different battery operating stages. Specifically, for different case studies (i.e., different battery operating stages), both the single-stage and the corresponding multi-stage battery degradation models are implemented in the scheduling framework as presented in Section \uppercase\expandafter{\romannumeral3} by replacing the battery cost penalty term with the corresponding model. Meanwhile, due to the superior performance of the multi-stage degradation model, the actual battery degradation with the single model is evaluated with the multi-stage model for each stage and utilized to calculate the actual operation cost. Assuming the battery operates in three stages for the three example days across a year, as shown in Fig. \ref{Flowchart}, the optimal scheduling strategies are obtained, which are compared respectively with the single degradation model counterparts. The optimization problem is implemented in Python with Pyomo as a modeling interface \cite{hart2011pyomo} and solved by IPOPT \cite{biegler2009large}.

The nominal energy of the battery is $E_{bat}$ = 1 MWh, and the battery operation range is within $SoC_{min}$ = 0 and $SoC_{max}$ = 1, the initial and ending SoC is 0.1, the round trip charge/discharge efficiency is set as $\eta_{ch}=\eta_{dis}$ = 0.92 and the cost coefficient $\lambda_k$ = 3560 DKK/kWh (478.55 EUR/kWh). The power limit $P_{max}$ is fixed to be 1 MW, and $P_{gmax}$ is set to be 10 MW. Assume the battery operation horizon is for one day (24-hour period) and the time interval is 30 min, thus $T$ = 48 and yields 96 decision variables. To compare the performance of the proposed multi-stage degradation model with the single-stage degradation model, numerical results corresponding to the optimal dispatch strategies for each case study are presented in Table \ref{comparison}. In addition, Fig. \ref{deg_com} presents the cumulative battery capacity degradation at different time points throughout the day.

\subsection{Case 1: Operating stage 1}
\begin{figure}
	\centering
	\begin{subfigure}{0.45\textwidth}
		\centering
		\includegraphics[width=\linewidth]{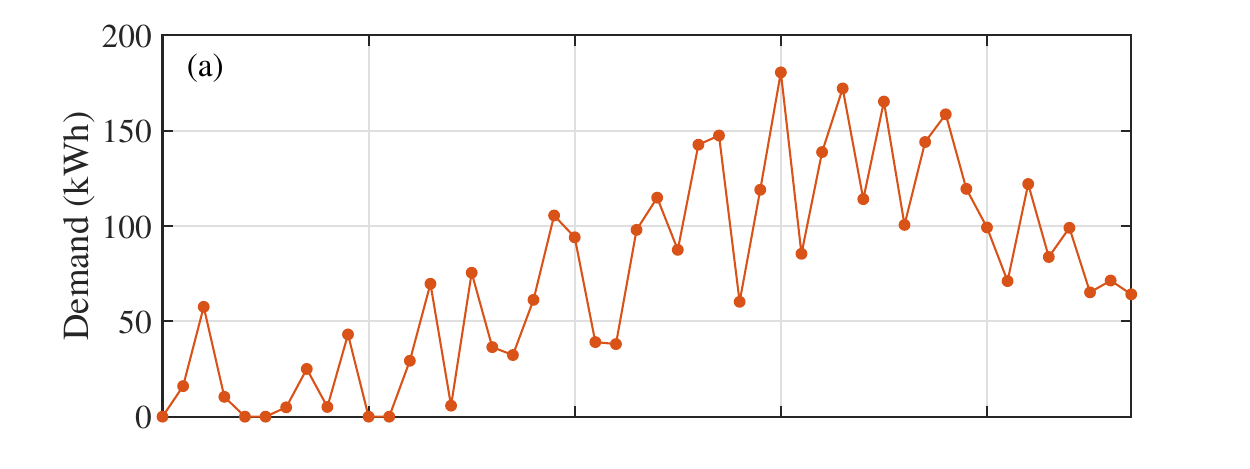}  
	\end{subfigure}
	
	\begin{subfigure}{0.45\textwidth}
	\vspace*{-0.255cm}
		\centering
		\includegraphics[width=\linewidth]{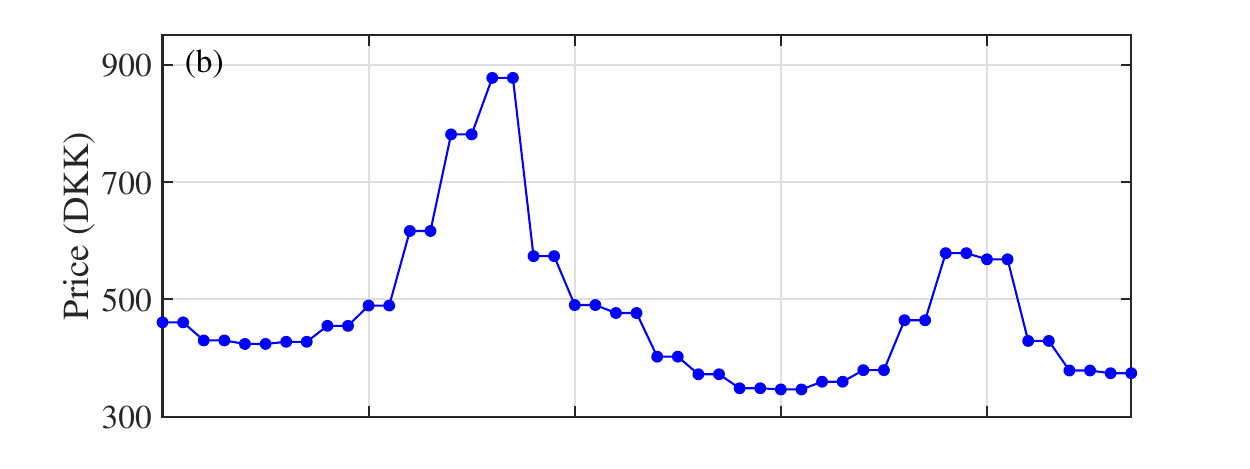}  
	\end{subfigure}
	
	\begin{subfigure}{0.45\textwidth}
	\vspace*{-0.255cm}
		\centering
		\includegraphics[width=\linewidth]{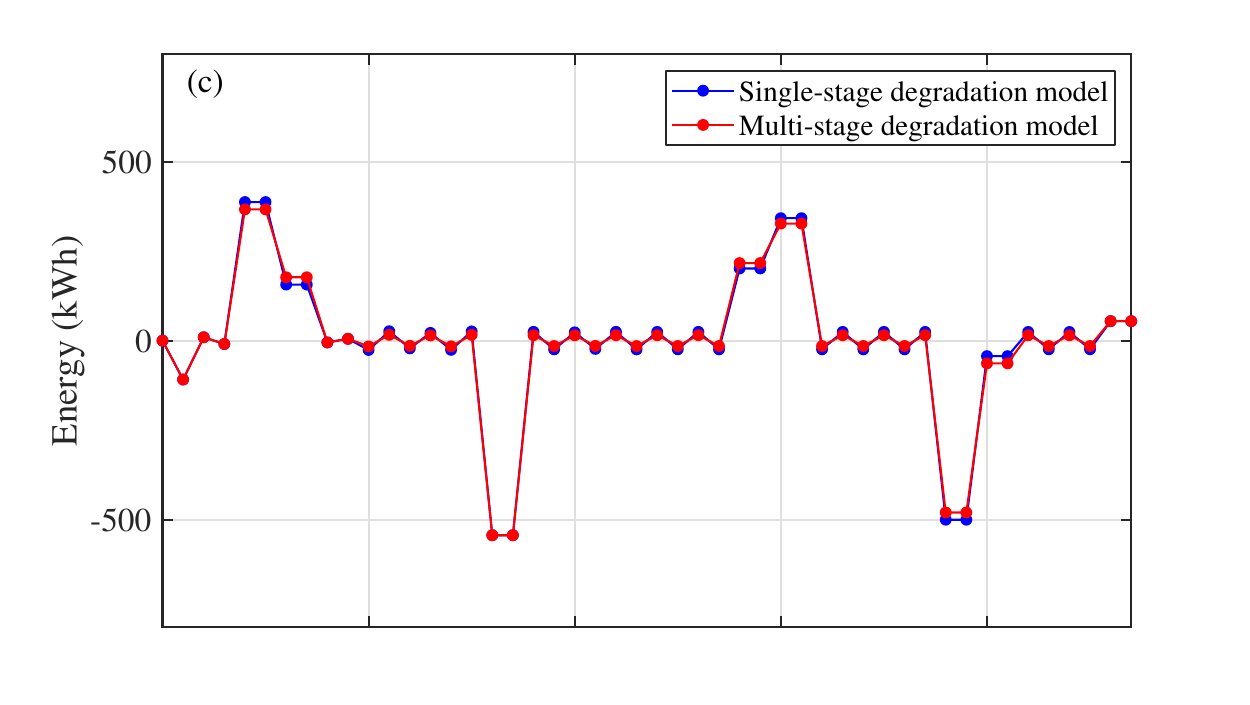}  
	\end{subfigure}
	
	\begin{subfigure}{0.45\textwidth}
	\vspace*{-0.46cm}
		\centering
		\includegraphics[width=\linewidth]{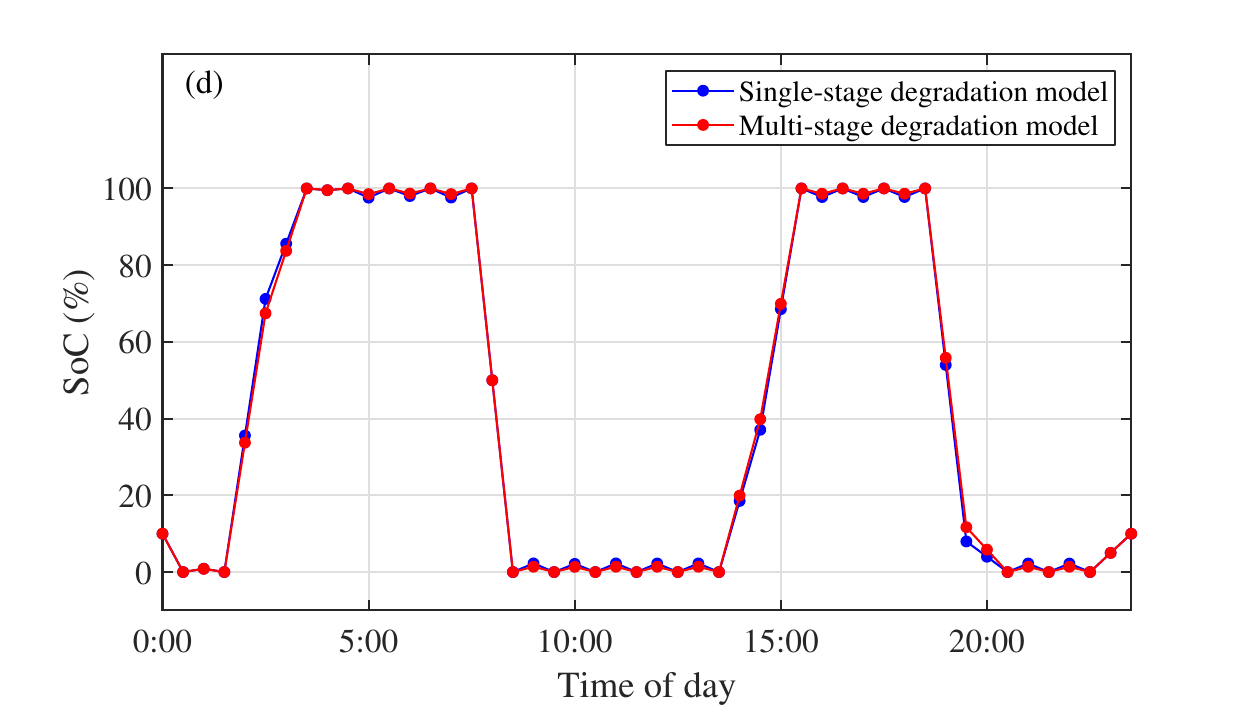}  
	\end{subfigure}	
	\caption{{Optimal dispatch strategies over a 24-hour period in stage 1. From top to bottom, the panels show (a) Load profile, (b) Spot market prices, (c) Battery operation profile, and (d) Battery SoC level.}}
	\label{profile_stage1}
\end{figure}

Assume the battery operating point of the selected day is in stage 1, where the degradation model parameters for the stage are shown in Table \uppercase\expandafter{\romannumeral3}. The load demand and the electricity spot prices for the selected day are shown in Fig. \ref{profile_stage1}(a)-(b). The battery operation profile (i.e., optimal dispatch strategies) for the two models is shown in Fig. \ref{profile_stage1}(c). In this case, battery charging operations are most desirable at around 2:00 and 15:00 when the electricity prices are the lowest, and discharging operations are mostly performed at around 8:00 and 19:00 when the electricity prices are the highest. It indicates that the optimal dispatch strategies for the battery operation are to keep the battery charged during a low-price period and discharged to supply energy to the load during a high-price period. In this way, by shifting the load consumption away from the peak prices, the overall operation cost could be reduced, and the impact of the oscillating load demand could also be mitigated.

Despite the slightly higher charging/discharging power command for the single degradation model compared with the multi-stage degradation model, it can be observed that the optimal operation strategies, as well as the battery SoC shown in the bottom subfigure with the two degradation models, are quite similar. Fig. \ref{deg1} shows that the accumulated battery capacity loss at the end of the day with the single degradation model is slightly larger than the multi-stage degradation model. In addition, based on the numerical results shown in Table \ref{comparison}, the energy arbitrage revenue with the single-stage degradation model is slightly higher than the multi-stage degradation model counterparts, while at the cost of more battery capacity losses, resulting in higher overall operation costs. In summary, the results above align with the model performance comparison for the first stage, where the battery degradation estimation accuracy for the two models is close.

\subsection{Case 2: Operating stage 2}
\begin{figure}
	\centering
	\begin{subfigure}{0.45\textwidth}
		\centering
		\includegraphics[width=\linewidth]{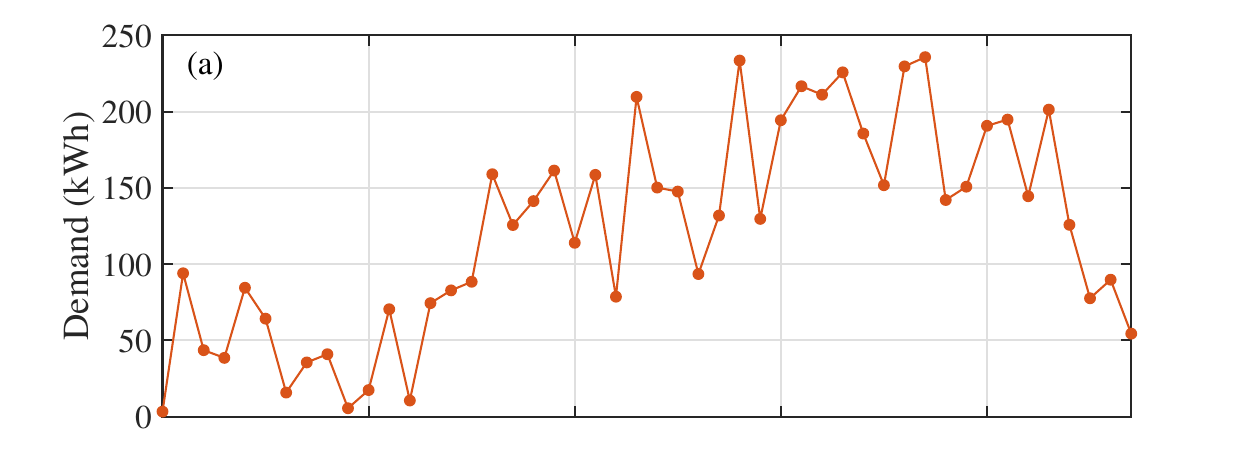}  
	\end{subfigure}
	
	\begin{subfigure}{0.45\textwidth}
	\vspace*{-0.255cm}
		\centering
		\includegraphics[width=\linewidth]{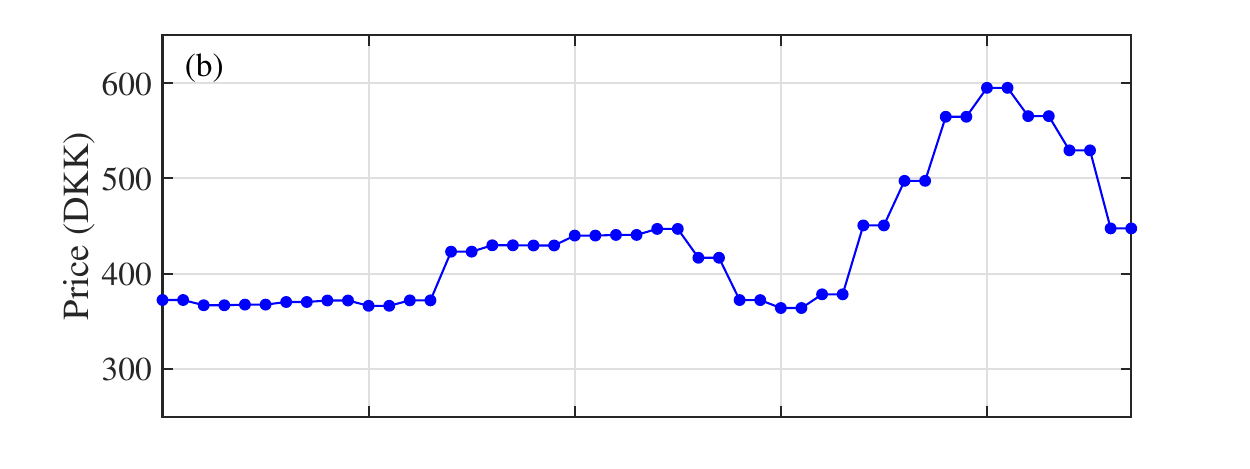}  
	\end{subfigure}
	
	\begin{subfigure}{0.45\textwidth}
	\vspace*{-0.255cm}
		\centering
		\includegraphics[width=\linewidth]{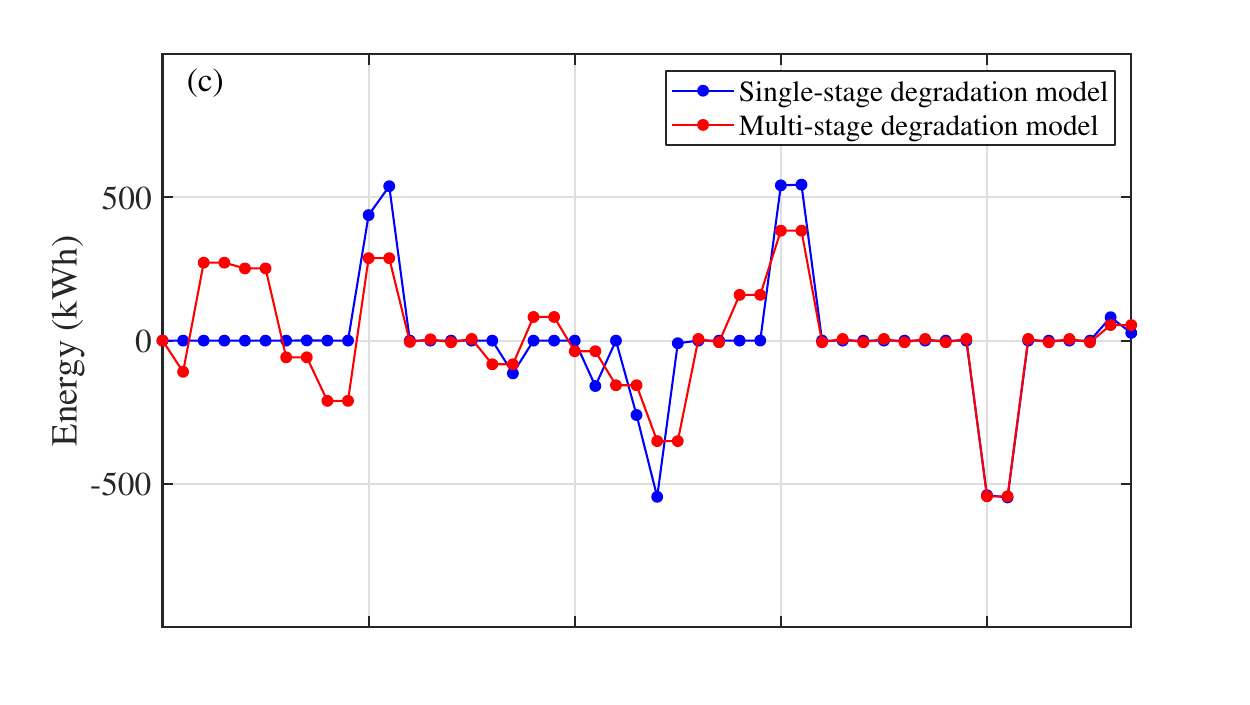}  
	\end{subfigure}
	
	\begin{subfigure}{0.45\textwidth}
	\vspace*{-0.46cm}
		\centering
		\includegraphics[width=\linewidth]{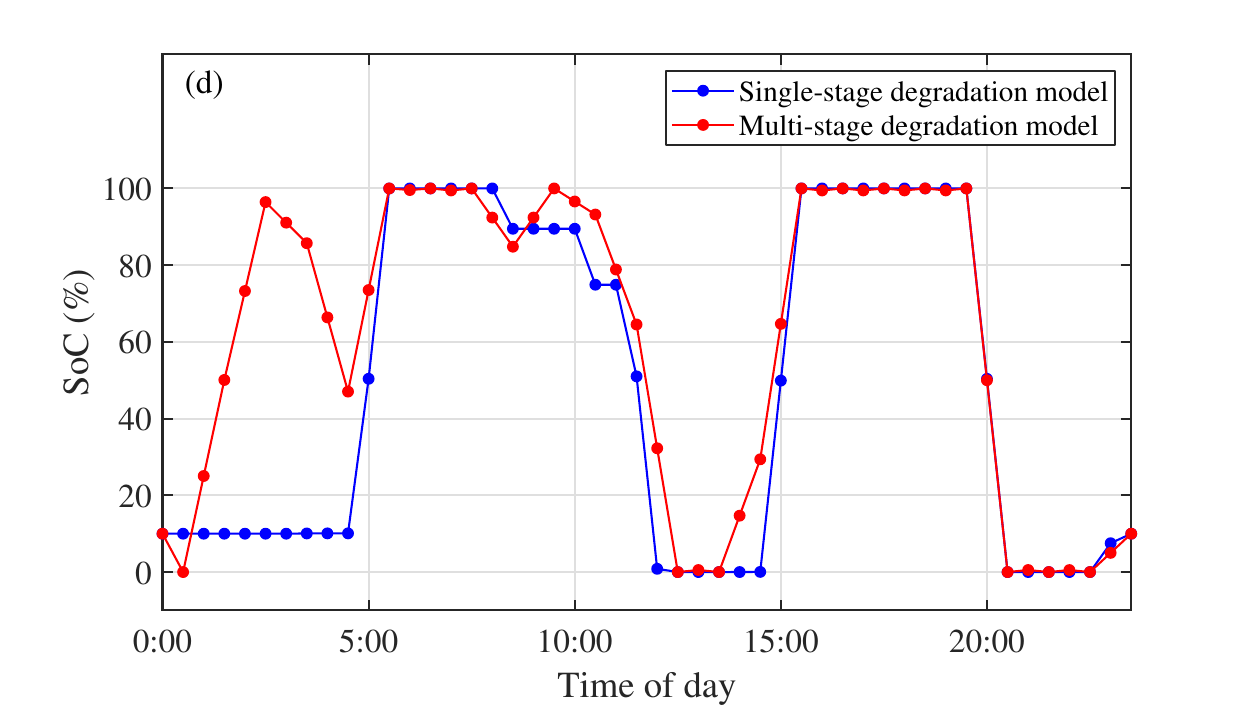}  
	\end{subfigure}	
	\caption{{Optimal dispatch strategies over a 24-hour period in stage 2. From top to bottom, the panels show (a) Load profile, (b) Spot market prices, (c) Battery operation profile, and (d) Battery SoC level.}}
	\label{profile_stage2}
\end{figure}

When the battery operates in stage 2, the model parameters in Table \uppercase\expandafter{\romannumeral4} are utilized. The load demand and the spot market prices for the example day in stage 2 are shown in Fig. \ref{profile_stage2}(a) and Fig. \ref{profile_stage2}(b), respectively. The battery operation profile and the corresponding SoC profile with the two degradation models are shown in the bottom two subfigures, where the optimal dispatch strategies are quite different for the two models. For both models, the battery is charged at around 5:00 and 15:00 during low price periods and discharged for power supply at around 8:00 and 20:00 during peak price periods. It could also be observed that the battery remains uncharged rather than supplying energy to the load when the load demand is high and oscillates continuously due to the increasing electricity prices during 16:00–19:30. It indicates a higher degradation cost for the period compared to the energy arbitrage revenue. While for the multi-stage model, the battery also operates during 0:00–5:00 when the electricity price is low. 

The resultant battery capacity losses process in Fig. \ref{deg2} shows that the battery capacity losses with the single-stage degradation model are larger than the multi-stage degradation model counterpart. Based on the numerical results of case 2 in Table \ref{comparison}, compared with the single-stage battery degradation model, the scheduling with the multi-stage degradation model is more desirable and results in around 3.3$\%$ reduction in the overall operation cost, which are 2053.21 and 1986.87 DKK, respectively, for the two models. 

\subsection{Case 3: Operating stage 3}
\begin{figure}
	\centering
	\begin{subfigure}{0.45\textwidth}
		\includegraphics[width=\linewidth]{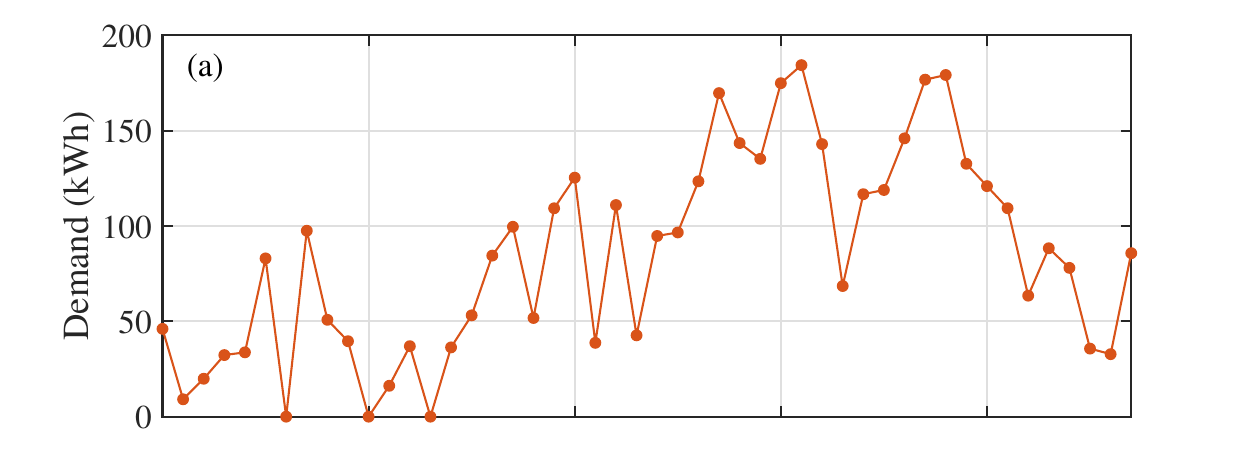}  
	\end{subfigure}
	
	\begin{subfigure}{0.45\textwidth}
	\vspace*{-0.255cm}
		\includegraphics[width=\textwidth]{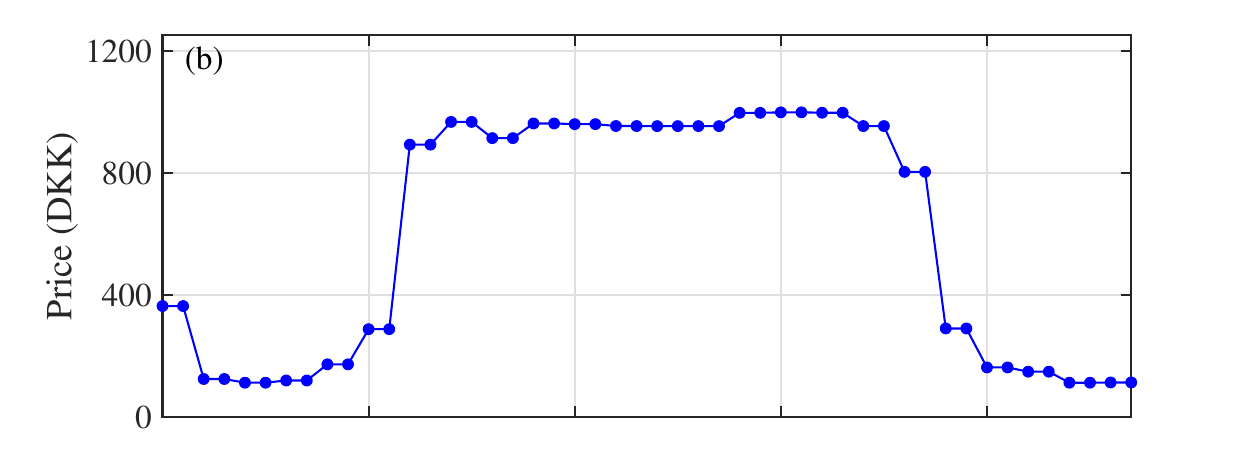}  
	\end{subfigure}
	
	\begin{subfigure}{0.45\textwidth}
	\vspace*{-0.255cm}
		\centering
		\includegraphics[width=\linewidth]{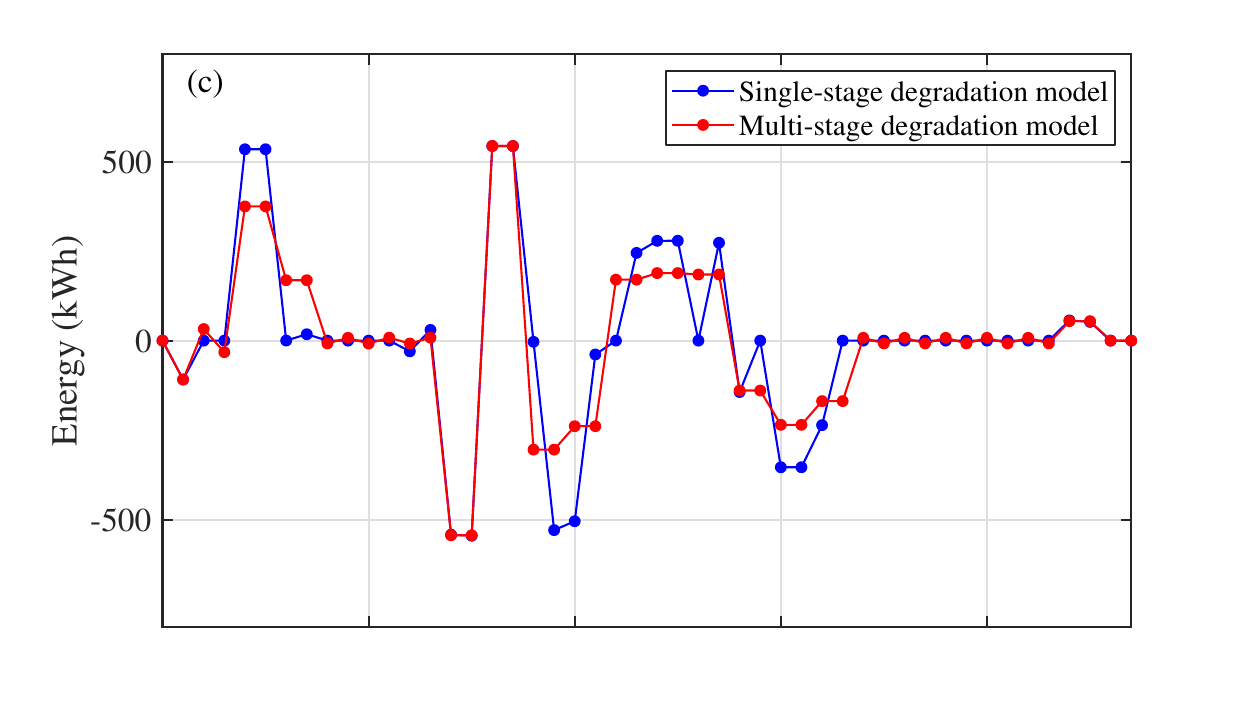}  
	\end{subfigure}
	
	\begin{subfigure}{0.45\textwidth}
	\vspace*{-0.46cm}
		\centering
		\includegraphics[width=\linewidth]{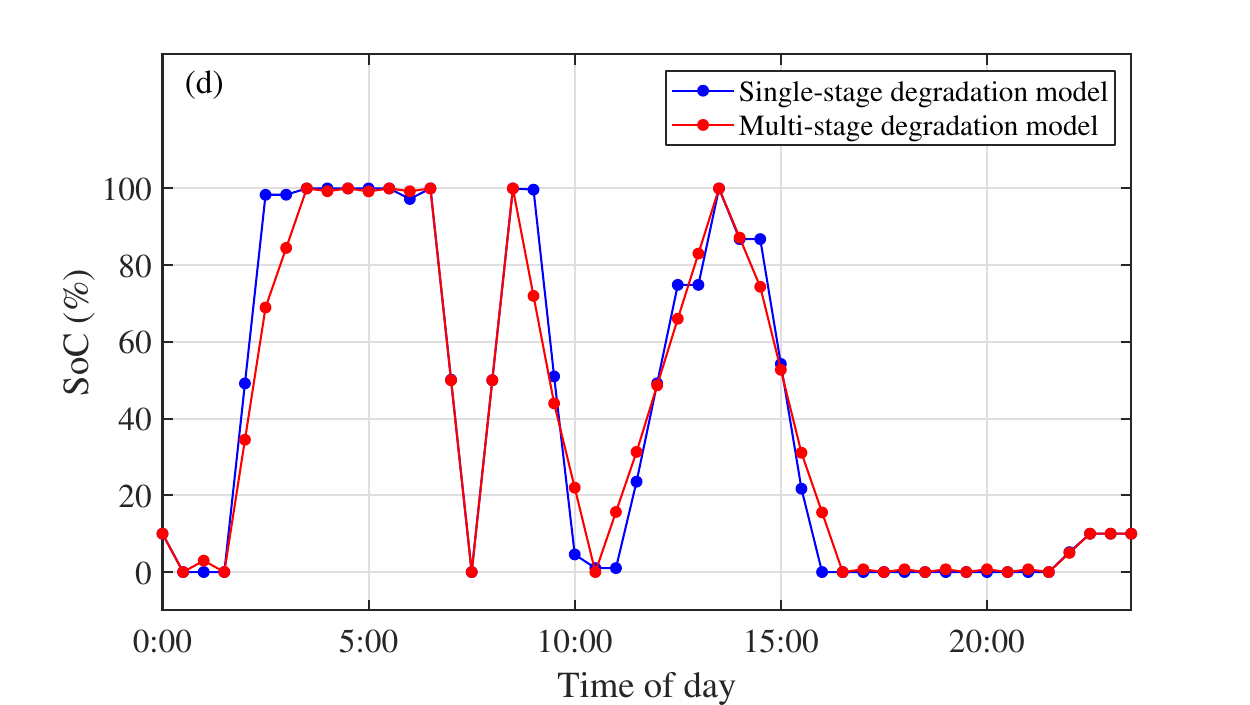}  
	\end{subfigure}	
	\caption{{Optimal dispatch strategies over a 24-hour period in stage 3. From top to bottom, the panels show (a) Load profile, (b) Spot market prices, (c) Battery operation profile, and (d) Battery SoC level.}}
	\label{profile_stage3}
\end{figure}

\begin{figure*}
	\centering
	\begin{subfigure}{0.32\textwidth}
		\centering
		\includegraphics[width=\linewidth]{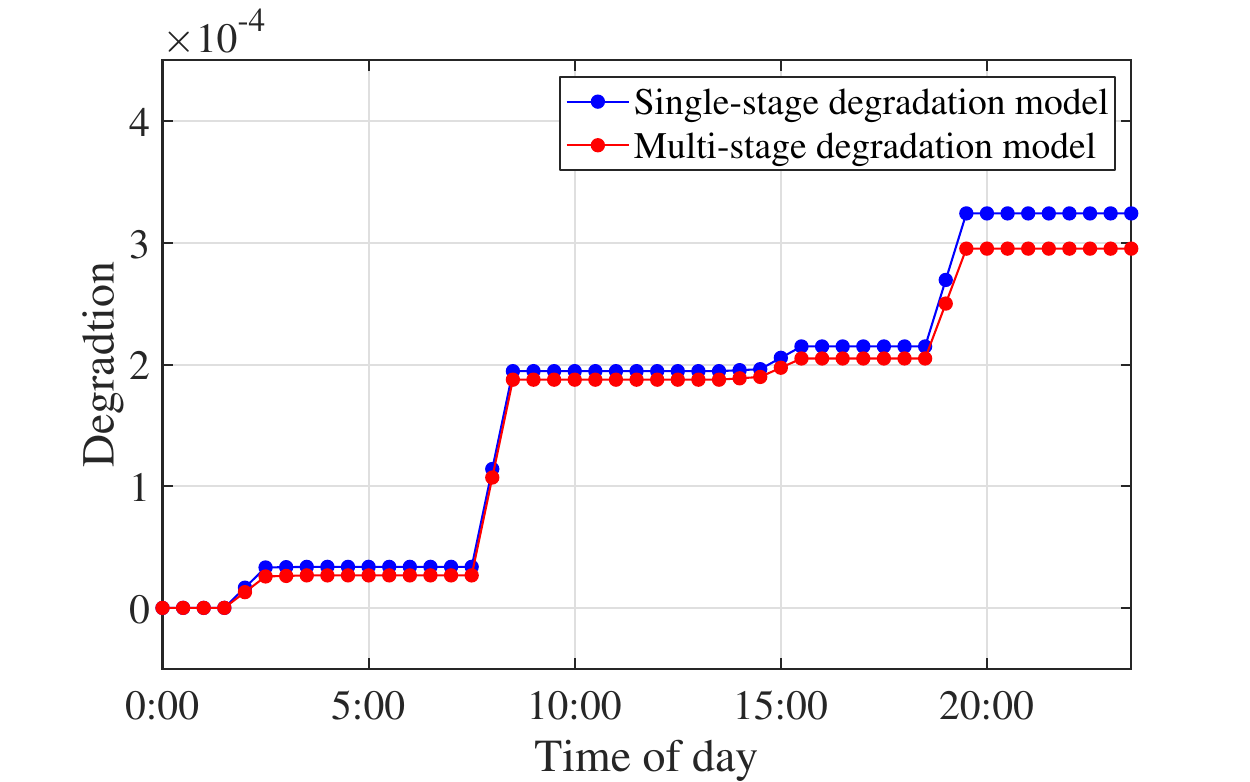}  
		\caption{Stage 1}
		\label{deg1}
	\end{subfigure}	
	\begin{subfigure}{0.32\textwidth}
		\centering
		\includegraphics[width=\linewidth]{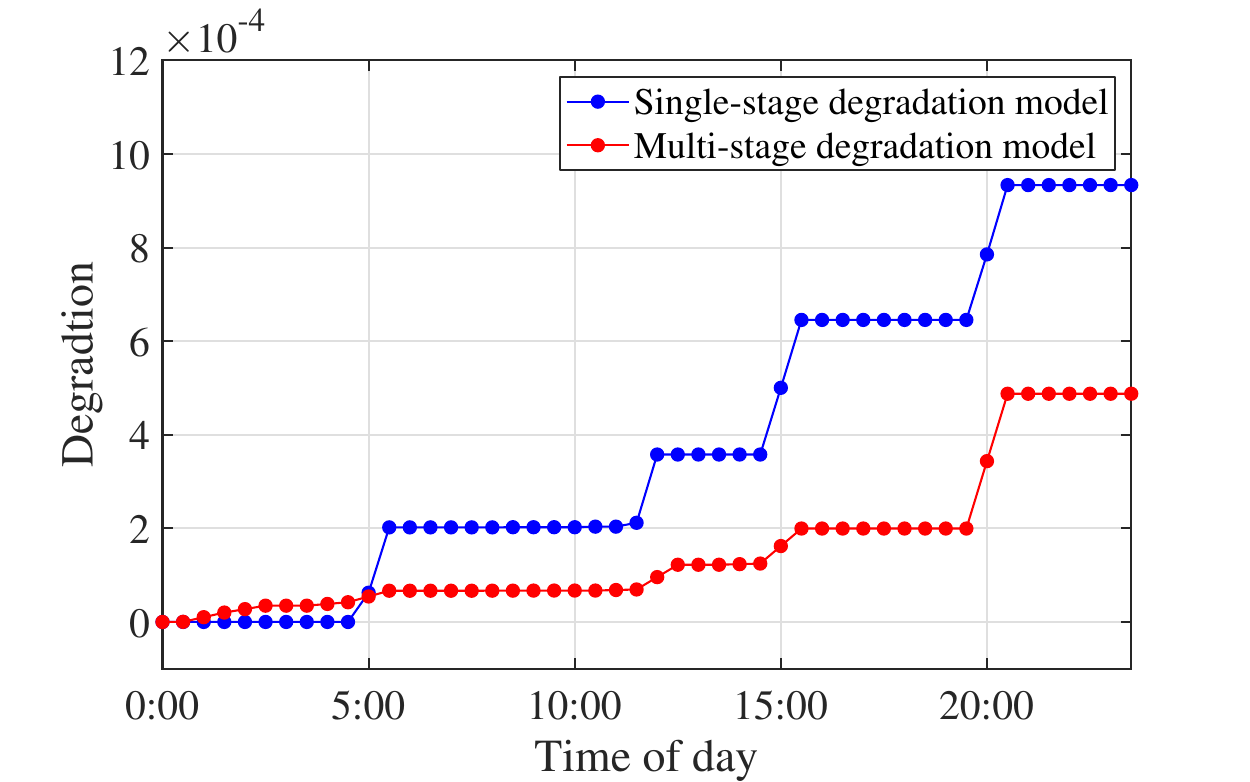}  
		\caption{Stage 2}
		\label{deg2}
	\end{subfigure}	
	\begin{subfigure}{0.32\textwidth}
		\centering
		\includegraphics[width=\linewidth]{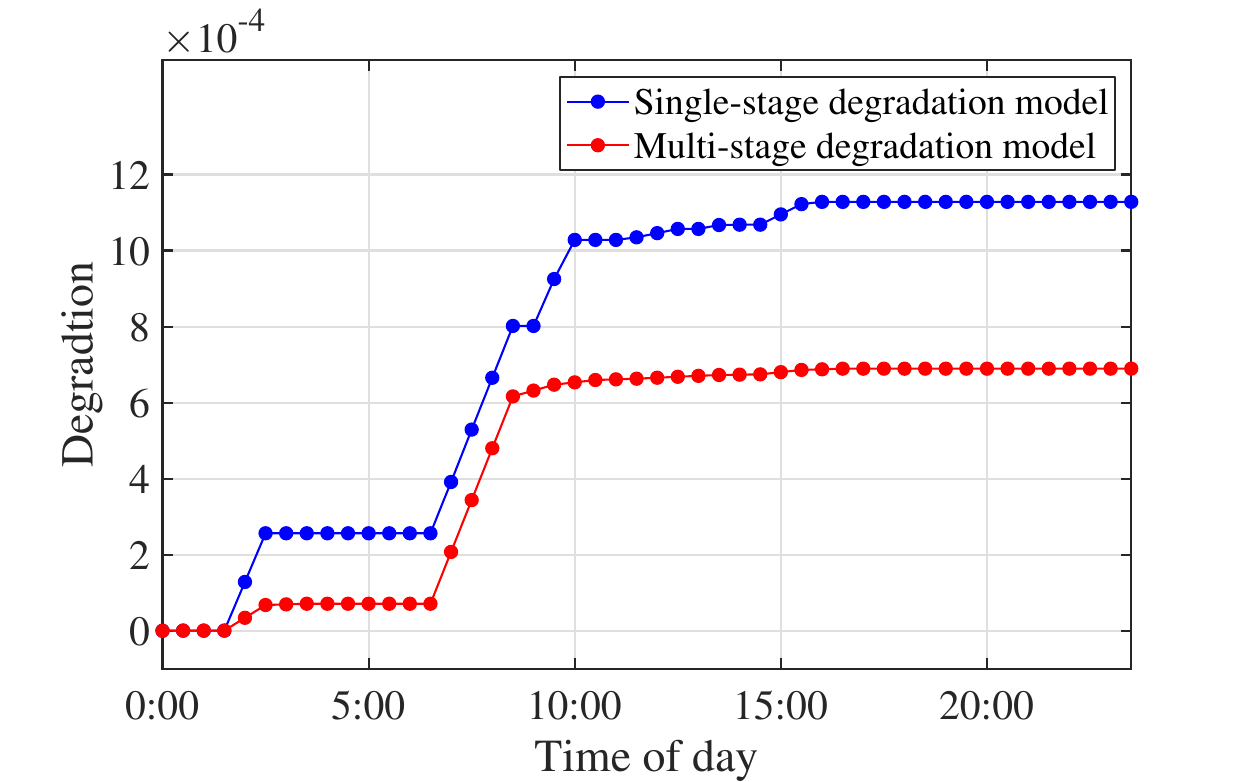}  
		\caption{Stage 3}
		\label{deg3}
	\end{subfigure}	
	\caption{{Comparison of the cumulative battery degradation for two models in each stage.}}
	\label{deg_com}
\end{figure*}

In stage 3, the model parameters in Table \uppercase\expandafter{\romannumeral5} are employed. The operation results regarding the load demand and electricity market price are shown in Fig. \ref{profile_stage3}(a)-(b). As the electricity price keeps at a high level during the daytime between 6:00 and 17:00, the battery is charged at around 2:00 when the price is the lowest and around 8:00 when the price drops. The battery is then discharged at around 6:00 when the price is at a high level, and around 8:30 when the price increases. Compared with the single-stage degradation model, the amplitude of charging and discharging power commands for the multi-stage model is reduced, resulting in reduced battery degradation, as shown in Fig. \ref{deg3}.

According to the numerical results in Table \ref{comparison}, the scheduling strategies with the single degradation model cause more capacity losses, though they achieve slightly higher energy arbitrage revenue than the multi-stage degradation model counterpart. As a result, the overall operation cost for scheduling with the multi-stage model is around 4.6$\%$ lower than the single-stage model counterpart.

\subsection{Discussion on the number of divided operating stages for battery throughout lifetime}
In principle, with the increasing number of divided stages over the battery lifetime, the accuracy of the multi-stage battery degradation model improves, and more desirable battery scheduling strategies could be obtained with the proposed optimization framework. To investigate the influence of the number of divided stages on the scheduling strategies, in the same way presented in Section \uppercase\expandafter{\romannumeral2}, by separating the battery degradation into different stages, the different multi-stage degradation models are implemented into the scheduling model.

\begin{figure}[H]
	\begin{center}
		\includegraphics[width=0.8\linewidth]{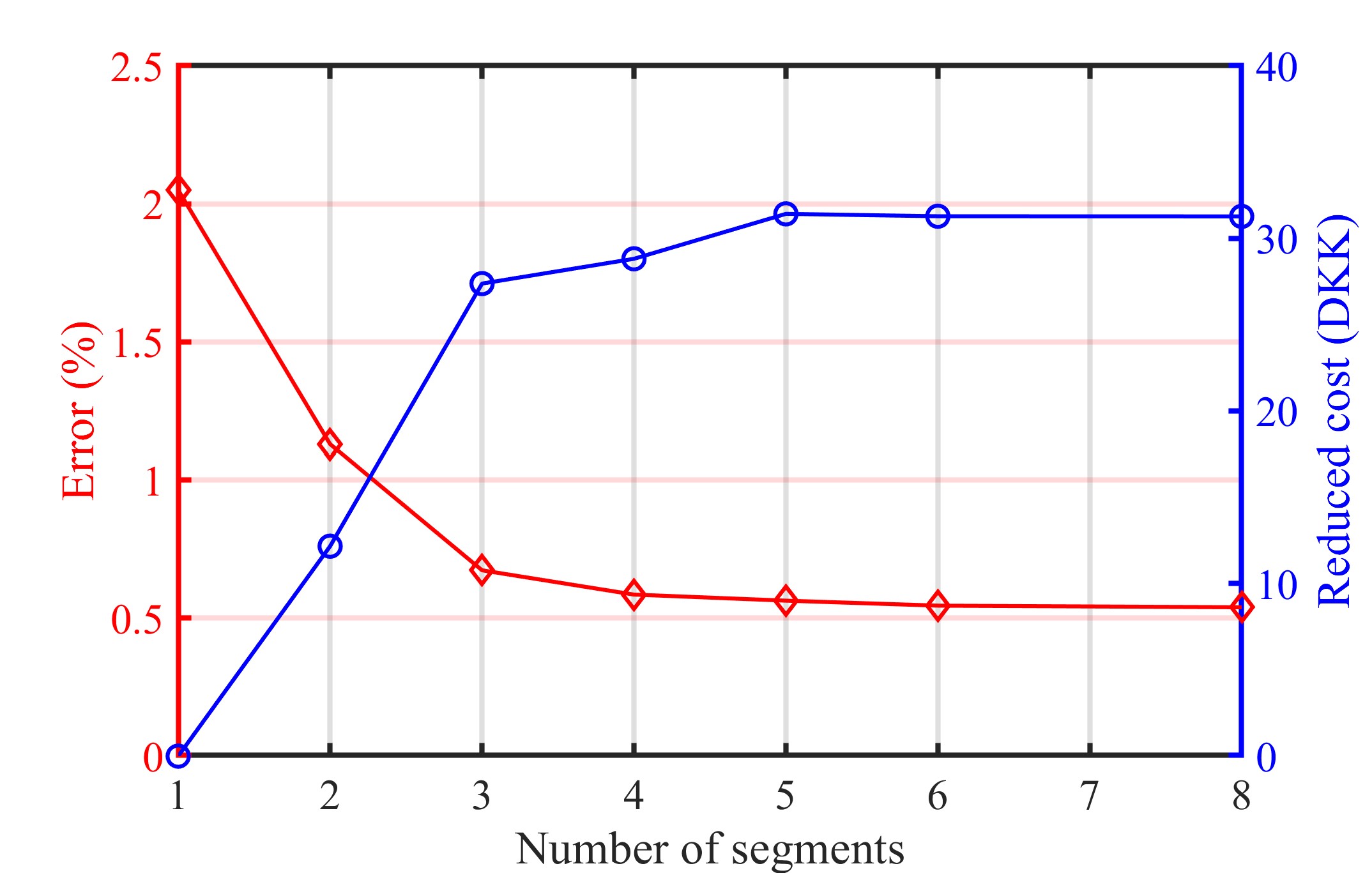}\\
		\caption{Model accuracy and average reduced operation cost using the proposed optimization framework based on multi-stage degradation model regarding different number of segments for battery's lifetime. (Blue line: Reduced cost; Red line: Model error.)}
		\label{number}
	\end{center}
\end{figure}

To demonstrate the scheduling results of different multi-stage degradation models, the reduced operation cost for each model compared with the results of the single-stage model is calculated. For simplification, the single-stage model and the different multi-stage models are both implemented for the FCS operational scheduling framework on a single day, irrespective of the different divided stages. The results of the average reduced operation cost of the single day for the multi-stage degradation models with the different number of divided stages are shown in Fig. \ref{number}.
As shown in Fig. \ref{number}, the improvement in model accuracy becomes marginal with the increasing number of divided stages. At the same time, it can be observed that the increase in reduced operation cost becomes stable at around 30 DKK on average after around five divided stages, at which point there are no evident benefits to further considering more divided stages.  

\subsection{Application of the multi-stage framework considering other battery degradation models}
To further validate the performance of the proposed multi-stage modeling method, we also applied the adaptive optimization framework with two battery degradation models in \cite{xu2017factoring} and \cite{gao2017lithium}. The model in \cite{xu2017factoring} is to demonstrate the applicability of our proposed multi-stage modeling method in another conventional model for the scheduling framework. To identify the parameters of the model, the battery cycling test data at different DoD levels are employed, and the parameters for different models are listed in Table \ref{model2_parameter}. Details of the parameterization method can be found in \cite{laresgoiti2015modeling}. Meanwhile, we also compare our method with the conventional multi-stage degradation model in \cite{gao2017lithium} by applying it to the proposed scheduling framework. To parameterize the model, the battery aging tests at different C-rate are employed. The models and corresponding parameters are listed in Table \ref{model2_parameter}.

\bgroup
\def\arraystretch{2.5}
\begin{table}[h]
\captionsetup{font={footnotesize}}
\caption{Model parameterization\label{model2_parameter}}
\centering
\resizebox{\linewidth}{!}{
\begin{tabular}{c c c}
\hline
Model& Single-stage model & Multi-stage models \\
\hline
$Q_{loss}=a*DoD^b$ & a=3.817e-4, b=3.226& \makecell{Stage 1: a=1.989e-3, b=4.226\\Stage 2: a=4.072e-4, b=3.497\\Stage 3: a=2.002e-4, b=2.788} \\[2ex]
\hline
$Q_{loss}=a*I^b+c$ & $\backslash$ & \makecell{Stage 1: a=2.36e-5, b=3.119, c=5.569e-5\\Stage 2: a=6.954e-6, b=3.092, c=1.348e-5\\Stage 3: a=3.778e-6, b=3.077, c=6.938e-6} \\
\hline
\end{tabular}}
\end{table}
\egroup

\bgroup
\def\arraystretch{1.8}
\begin{table*}[b]
\caption{Comparison of the optimal scheduling results with different degradation models.\label{comparison}}
\centering
\resizebox{\linewidth}{!}{
\begin{tabular}{c c c c c c c c c c}
\hline
\multirow{2}{*}{Cases} & \multicolumn{3}{c}{Case 1} & \multicolumn{3}{c}{Case 2} & \multicolumn{3}{c}{Case 3}\\
\cline{2-10}
& \cite{gao2017lithium}& Single-stage model & Multi-stage model & \cite{gao2017lithium} & Single-stage model & Multi-stage model& \cite{gao2017lithium} & Single-stage model & Multi-stage model \\
\hline
\makecell[c]{Battery degradation \\ ($\%$)} & 0.0483& 0.0324 & 0.0295 & 0.155& 0.0934 & 0.0488 & 0.260& 0.113 & 0.0689\\

\makecell[c]{Energy arbitrage revenue \\ (DKK)} & 1215.02& 1212.96 & 1212.33 &708.34 &697.66 & 699.79 & 1420.30&1418.68 & 1416.21\\

\makecell[c]{Operation cost \\ (DKK)}  & 484.27& 450.46 & 446.92 & 2173.63&2053.21 & 1986.87 & 1646.28 & 1559.60 & 1487.09\\
\hline
\end{tabular}}
\end{table*}
\egroup

As mentioned, the multi-stage modeling method in \cite{gao2017lithium} is not explored for the scheduling problem, and it considers a single current stress factor, which also lacks the flexibility of incorporating multiple factors into the degradation model. The scheduling results with the multi-stage modeling method are added to Table \ref{comparison} for better comparison. It can be observed that the conventional multi-stage model underestimates the battery degradation at different stages. Thus, the battery usage is increased, resulting in higher energy arbitrage revenue while also inducing higher overall operation cost compared with the proposed multi-stage modeling method counterpart.

\bgroup
\def\arraystretch{1.8}
\begin{table*}
\caption{Performance of the proposed multi-stage degradation modeling method based on battery degradation model in \cite{xu2017factoring}.\label{comparison_model2}}
\centering
\resizebox{\linewidth}{!}{
\begin{tabular}{ c  c  c  c  c  c  c }
\hline
\multirow{2}{*}{Cases} & \multicolumn{2}{c}{Case 1} & \multicolumn{2}{c}{Case 2} & \multicolumn{2}{c}{Case 3}\\
\cline{2-7}
& Single-stage model & Multi-stage model & Single-stage model & Multi-stage model & Single-stage model & Multi-stage model \\
\hline
\makecell[c]{Battery degradation \\ ($\%$)} & 0.0563 & 0.0365 & 0.0369 & 0.0340 & 0.0261 & 0.0287\\

\makecell[c]{Energy arbitrage revenue \\ (DKK)} & 1214.43 & 1211.40 & 697.64 & 702.82 & 1338.33 & 1419.67\\

\makecell[c]{Operation cost \\ (DKK)}  & 498.43 & 467.73 & 1981.62 & 1971.67 & 1457.38 & 1406.65\\
\hline
\end{tabular}}
\end{table*}
\egroup

In addition, we also demonstrate the applicability of the proposed multi-stage modeling method on another battery degradation model from \cite{xu2017factoring}. The multi-stage models are implemented into the scheduling model in the same way presented in Fig. \ref{Flowchart}, and the quantitative comparison between the single battery degradation model and the derived multi-stage models at different stages is shown in Table \ref{comparison_model2}.

It can be inferred from the model parameters that the battery degrading pattern varies at different stages, and the battery degrades faster with higher DoD at stage 1 while stabilizing at stages 2 and 3. At stage 1, the single-stage model underestimates the battery degradation, and even though higher energy arbitrage revenue is obtained while the overall operation cost is higher compared with the multi-stage model counterpart. While for the rest two stages, the single-stage model overestimates the battery degradation. Thus, the energy arbitrage revenue is reduced, and the overall operation cost is higher compared with the scheduling results of multi-stage model counterparts. The operation cost of the proposed method is reduced by 3.4\% on average compared with the single-stage model counterpart.

In summary, the proposed multi-stage degradation modeling method is independent of different conventional battery degradation models, which could be employed to capture the varying degradation patterns to project the battery degradation across its whole lifetime more accurately, thus obtaining optimal scheduling strategies for minimizing the operation cost.\\

\subsection{Discussion on the practical implementation}
The average computation time of the proposed optimization framework for the case study is only around 5.02 s, carried out on a PC with an Intel (R) Core (TM) i5-10210U 1.60 GHz processor and 8.00 GB of RAM. To further investigate the computation time of the formulated problem, the optimization problem is solved under various conditions. Moreover, the average computation time to obtain the optimal charging and discharging decisions is around 5.65 s. As the scheduling problem is formulated for the day-ahead stage, the computation time is more than acceptable for day-ahead optimization, where updates {red}{of} schedule will typically be calculated once every several minutes or more. Furthermore, suppose more ancillary services are incorporated into the optimization problem, such as frequency regulation, which has a smaller time step. In that case, the scheduling model should be properly formulated to satisfy the computation time for the specific service.

In addition, the proposed multi-stage degradation modeling method is applicable for other practical systems equipped with BESS by improving the accuracy of assessing the battery degradation across its entire lifetime. Moreover, the scheduling with the multi-stage models can also be deployed as part of the cloud-based real-time control platform presented in \cite{Gebbran2022cloud}.

In summary, the proposed optimization framework with the multi-stage battery degradation model could be implemented in practical applications.

\section{Conclusion}
In this paper, a novel adaptive optimization framework with a multi-stage battery degradation modeling method is proposed. The multi-stage degradation modeling method is proposed to capture the varying battery aging patterns across its entire lifetime, thus achieving an accurate evaluation of the aging cost for the scheduling problems in different battery operational stages. Case studies for the scheduling problem at different stages by applying the multi-stage modeling method to a generic cycle degradation model, resulting in around 2.9$\%$ reduction in the overall operation cost compared with the single-stage model counterpart. Additionally, as the number of segments for battery lifespan increases, the model error decreases and stabilizes at roughly 0.5$\%$ with enough separated segments for battery degradation, while the reduced operation cost increases and saturates. The superiority of the proposed multi-stage framework is further validated based on other battery degradation models. The computation time for the day-ahead scheduling problem is around 5 s, which makes the proposed optimization framework practical for real-time applications. 

More importantly, the proposed multi-stage degradation modeling method is independent of various battery chemistry types and empirical/semi-empirical degradation models. Moreover, for other systems equipped with BESSs, by properly formulating the schedule model, the proposed optimization framework with the multi-stage degradation model could still be employed to obtain the economic operations. Results in the study provide a guide for battery scheduling in different SoH stages, concluding that the multi-stage modeling method in the proposed optimization framework achieves more cost-effective dispatch strategies. Future work could be mitigating the requirement of the experimental aging test across the whole lifespan for parameterizing the multi-stage model, which is time-consuming. Furthermore, other external factors, such as temperature, impact the battery degradation rate. While including them in operational planning problems imbues the modeling framework with additional complexity and uncertainty (and hence, out of the scope of this manuscript), they may prove to improve the end-cost of operation further when accounting for battery degradation. Therefore, the authors further highlight the inclusion of other external degradation factors as future research topics.

\bibliographystyle{IEEEtran}
\bibliography{IEEEabrv,Bibliography}

\end{document}